\documentclass[sigconf]{acmart}

\usepackage{epsfig}
\usepackage{graphicx}
\usepackage{amsmath}
\usepackage{amssymb}

\usepackage{bm}
\usepackage{enumitem}
\usepackage{tablefootnote}
\usepackage[linesnumbered,titlenumbered,ruled,vlined,commentsnumbered,noend]{algorithm2e}
\usepackage{stfloats}
\usepackage{makecell}
\usepackage{colortbl}
\definecolor{linen}{rgb}{0.96, 0.94, 0.93}
\definecolor{lightcyan}{rgb}{0.88, 1.0, 1.0}
\definecolor{lightyellow}{rgb}{1.0, 1.0, 0.88}
\definecolor{azure}{rgb}{0.94, 1.0, 1.0}

\usepackage[font=footnotesize,labelfont=bf]{caption}

\usepackage{multirow}
\usepackage{tabularx}
\usepackage{dsfont}
\usepackage{url}
\usepackage[tight]{subfigure}
\usepackage{color}
\usepackage{subfigure}
\usepackage{amsthm}
\usepackage[export]{adjustbox}
\usepackage{comment}

\usepackage[utf8]{inputenc} 
\usepackage[T1]{fontenc}    
\usepackage{url}            
\usepackage{booktabs}       
\usepackage{amsfonts}       
\usepackage{nicefrac}       
\usepackage{microtype}      
\usepackage{sidecap}

\definecolor{dg}{rgb}{0,0.694,0.298}
\definecolor{purple}{rgb}{0.4,0.176,0.569}

\usepackage{pifont}
\usepackage{xspace}
\makeatletter
\DeclareRobustCommand\onedot{\futurelet\@let@token\@onedot}
\def\@onedot{\ifx\@let@token.\else.\null\fi\xspace}
\def\eg{\emph{e.g}\onedot}

\def\etc{\emph{etc}\onedot} 
 
\def\etal{\emph{et al}\onedot}
\makeatother

\AtBeginDocument{%
  \providecommand\BibTeX{{%
    \normalfont B\kern-0.5em{\scshape i\kern-0.25em b}\kern-0.8em\TeX}}}

\setcopyright{acmcopyright}

\copyrightyear{2021} 
\acmYear{2021} 
\setcopyright{acmcopyright}\acmConference[MM '21]{Proceedings of the 29th ACM International Conference on Multimedia}{October 20--24, 2021}{Virtual Event, China}
\acmBooktitle{Proceedings of the 29th ACM International Conference on Multimedia (MM '21), October 20--24, 2021, Virtual Event, China}
\acmPrice{15.00}
\acmDOI{10.1145/3474085.3475518}
\acmISBN{978-1-4503-8651-7/21/10}

\begin{document}
\fancyhead{}

\title{\emph{FakeTagger}: Robust Safeguards against DeepFake Dissemination via Provenance Tracking}

\author{Run Wang$^{1,2,\dagger}$, \ Felix Juefei-Xu$^{3}$, \ Meng Luo$^{4}$, \ Yang Liu$^{5}$, \ Lina Wang$^{1,2}$}
\thanks{$^{\dagger}$ Run Wang is the corresponding author~({wangrun@whu.edu.cn})}


\affiliation{\institution{
$^{1}$School of Cyber Science and Engineering, Wuhan University, China \ \ }}
\affiliation{\institution{ $^{2}$Key Laboratory of Aerospace Information Security and Trusted Computing, Ministry of Education, China\\}}
\affiliation{\institution{$^{3}$Alibaba Group, USA     $^{4}$Northeastern University, USA     $^{5}$Nanyang Technological University, Singapore}}




\renewcommand{\shortauthors}{Run Wang et al.}

\begin{abstract}

In recent years, DeepFake is becoming a common threat to our society, due to the remarkable progress of generative adversarial networks (GAN) in image synthesis. Unfortunately, existing studies that propose various approaches, in fighting against DeepFake and determining if the facial image is real or fake, is still at an early stage. Obviously, the current DeepFake detection method struggles to catch the rapid progress of GANs, especially in the adversarial scenarios where attackers can evade the detection intentionally, such as adding perturbations to fool the DNN-based detectors. While passive detection simply tells whether the image is fake or real, DeepFake provenance, on the other hand, provides clues for tracking the sources in DeepFake forensics. Thus, the tracked fake images could be blocked immediately by administrators and avoid further spread in social networks.

In this paper, we investigate the potentials of image tagging in serving the DeepFake provenance tracking. Specifically, we devise a deep learning-based approach, named \emph{FakeTagger}, with a simple yet effective encoder and decoder design along with channel coding to embed message to the facial image,  which is to recover the embedded message after various \textbf{drastic} GAN-based DeepFake transformation with high confidence. The embedded message could be employed to represent the identity of facial images, which further contributed to DeepFake detection and provenance. Experimental results demonstrate that our proposed approach could recover the embedded message with an average accuracy of more than 95\% over the four common types of DeepFakes. Our research finding confirms effective privacy-preserving techniques for protecting personal photos from being DeepFaked. 

\begin{figure}[t]
\centering
\includegraphics[width=0.95\linewidth]{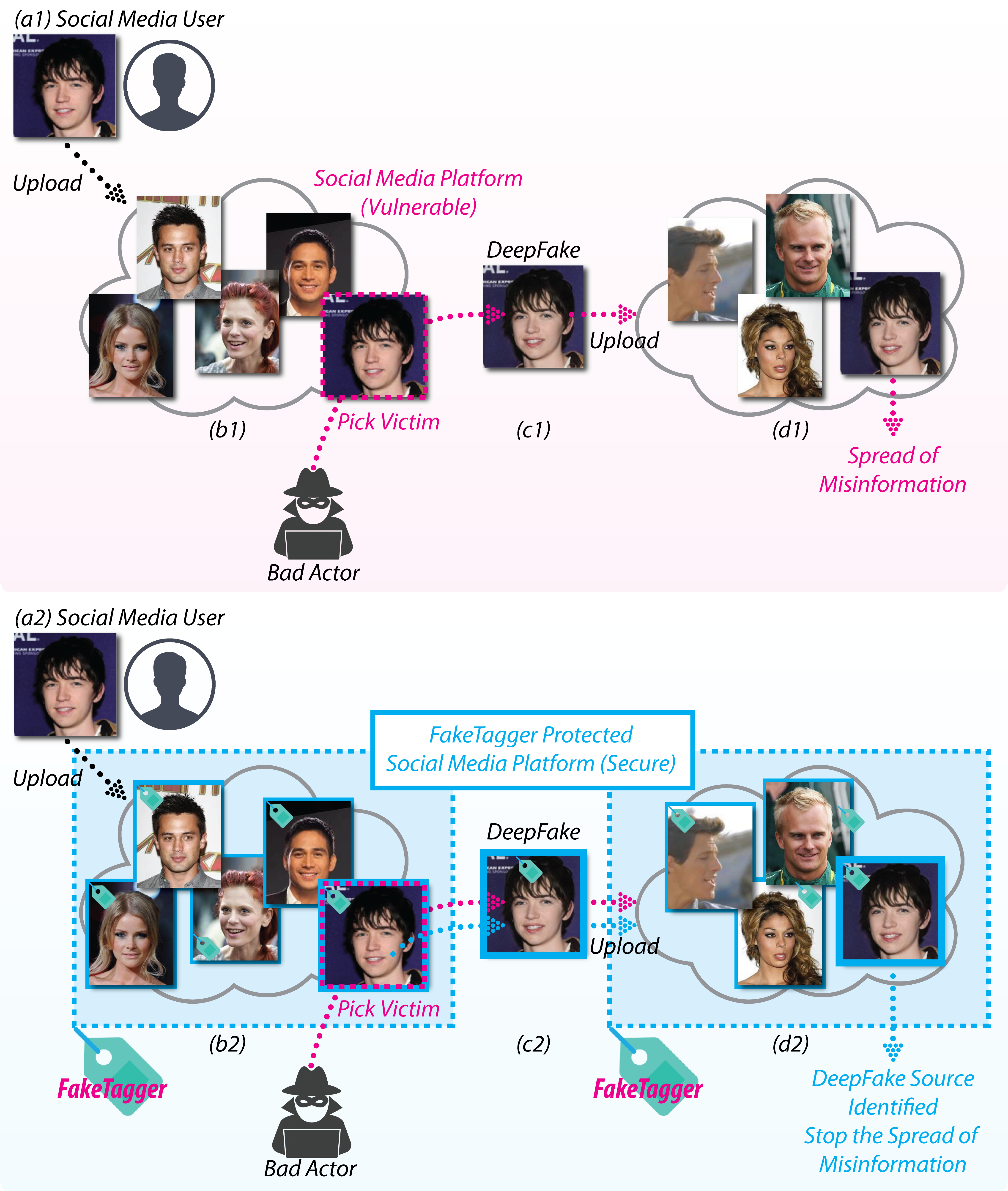}
\caption{Comparison between a vulnerable social media platform (top panel) and a FakeTagger protected social media platform (bottom panel) in handling malicious bad actors for spreading the misinformation by using DeepFake technology.}
\label{fig1}
\vspace{-10pt}
\end{figure}

\end{abstract}





\maketitle


\section{Introduction}\label{sec:intro}


Capturing the exciting moments with camera and sharing them with friends over social networks (\eg, Facebook, Twitter, Instagram) becomes a common activity in our daily life. However, with the recent rapid development of GAN and its variants in image synthesis, our shared personal photos may suffer from being manipulated by various GANs to create DeepFakes \cite{mirsky2020creation,juefei2021countering}. Abusing the DeepFakes can bring potential threats and concerns to everyone, for example, releasing a realistic fake statement \cite{fake-statement}, creating fake pornography \cite{porn-pornography}, \etc. Additionally, many freely available tools (\eg, FaceApp, ZAO) allow users to easily create DeepFakes on their own without any additional expertise. More importantly, the synthesized image/ videos are indistinguishable to our eyes and we are living in a world where we cannot believe our eyes anymore. Thus, effective measures should be developed for fighting against such DeepFakes to protect our personal security and privacy.

In the past two years, researchers are actively proposing various DeepFake detection techniques to determine if a suspicious still image or video is real or fake passively. These studies mostly focus on capturing the minor differences between real and synthesized images as the detection clues, for instance, examining the visible artifacts in the synthesized images \cite{li2018ictu,yang2019exposing}, investigating the invisible artifacts in the frequency domain \cite{frank2020leveraging,barni2020cnn}, and observing the unreplicable biological signals from real videos \cite{acmmm20_deeprhythm,ciftci2020fakecatcher}. Unfortunately, these approaches are not practical with poor performance in dealing with DeepFake created with unseen synthetic techniques and spread in the real word where suffers various degradations (\eg{}, compression, blurring, resizing ). According to the results of latest DeepFake detection competition (DFDC) hosted by Facebook, the best detection result gives less than 70\% accuracy in spotting DeepFakes in the real-world. In general, the existing DeepFake detection methods suffer the following two key challenges.

\begin{itemize}[leftmargin=*]
    \item \textbf{Poor generalization on unseen synthetic techniques}. Almost all the existing studies focus on evaluating the effectiveness of their method on a limited number of known GANs or simple datasets. Since advanced GANs will be developed at an enormous speed and the visible or invisible artifacts which could be employed in previous GANs for distinguishing real and fake will likely be removed or corrupted \cite{karras2020analyzing,choi2020stargan}.
    \item \textbf{Not robust against image quality degradations}. In the real-world scenario, DeepFakes suffer various degradations, including simple image transformation (\eg, resizing, compression, Gaussian noises) \cite{qian2020thinking,huang2020fakepolisher} and adversarial noise attack with imperceptible perturbations \cite{carlini2020evading,gandhi2020adversarial}, which is the biggest obstacle in developing robust DeepFake detectors.
\end{itemize}

To address the aforementioned two inevitable key challenges in DeepFake detection, recently, researchers approach the DeepFake defense proactively by adding imperceptible adversarial noises to disrupt the GAN-based image synthesis \cite{ruiz2020disrupting,yeh2020disrupting}, instead of merely improving the generation capabilities in unknown GANs and robustness against various degradations in detecting DeepFakes passively. However, in disrupting DeepFakes, the added adversarial perturbations could be easily figured out by detectors \cite{xu2017feature,metzen2017detecting} and the imperceptible adversarial perturbations are fragile which could be easily destroyed \cite{samangouei2018defense,guo2017countering}. In this paper, we propose a novel approach, named \emph{FakeTagger}, by protecting the safety and privacy of faces with image tagging to embed messages into the victim images and recover them after being DeepFaked to determine whether they are DeepFaked and manipulated by GANs proactively. Specifically, our proposed approach can be employed for DeepFake forensics for both detection and provenance purposes to track the source identity of DeepFakes.


Our FakeTagger is motivated by the provenance and tracking idea with sensible tags which is widely applied in protecting the safety of food in production and selling. Similarity, our personal image spread in the social network also need protection for tracing its illegally manipulation. In designing our FakeTagger, the following three key challenges should be well addressed. 1) \textbf{Tackling diverse GANs}. In creating DeepFakes, attackers could employ various GANs (\eg{}, entire synthesis, partial synthesis) with different architectures, while the employed GANs are unknown to us. Thus, our FakeTagger should recover embedded tags from images manipulated with unseen GANs. 2) \textbf{Robust against image transformations}. In the real-world scenario, the embedded images will suffer common image transformation after GAN-based manipulations, thus our FakeTagger should recover the embedded tag in such case. 3) \textbf{Stealthiness of embedded tags}. The embedded tags should insensitive to our human eyes.

To address the aforementioned challenges in embedding a message into the images, in this paper, our proposed FakeTagger is based on a simple yet effective encoder and decoder architecture by incorporating channel coding that could recover messages effectively even after drastic GAN-based transformation. The introduced channel coding is designed for injecting redundant messages to improve its robustness. In FakeTagger, a DeepFake simulator connects the encoder and decoder to simulate various manipulations with GANs on the encoded images to enforce that the decoder could recover the embedded messages effectively after GAN-based transformation. To comprehensively evaluate the effectiveness of our FakeTagger, our experiments are conducted on the existing four types of DeepFake, including \textbf{identity swap}, \textbf{face reenactment}, \textbf{attribute editing}, and \textbf{entire synthesis}. Experimental results have demonstrated that FakeTagger achieves an average accuracy of nearly 95\% on the four types of DeepFakes in recovering the embedded messages. Our main contribution are summarized as follows:

\begin{itemize}[leftmargin=*]
    \item \textbf{Hint new research direction in defending DeepFake with image tagging.} To the best of our knowledge, this is the first work proposing image tagging for DeepFake provenance and tracking. Our proactive defense techniques well address the generalization and robustness issues in the traditional artifact-based DeepFake detection. Our work opens a new research direction in defending DeepFakes towards tracking the source of DeepFakes for aiding forensics further. 
    \item \textbf{Presenting an effective method for image tagging.} We devise a simple yet effective method for image tagging by applying a jointly trained encoder and decoder for message embedding and recovering. We introduce channel coding to inject redundancy to improve its resistance on DeepFake transformation.
    \item \textbf{Performing a comprehensive evaluation on typical DeepFakes.} Experiments are conducted on four types of DeepFakes spanning identity swap, face reenactment, attribute editing, and entire synthesis. Experimental results demonstrated the effectiveness in recovering messages from drastic GAN-based transformation in both white-box and black-box settings.
    
\end{itemize}

\section{Related Work}\label{sec:related}



\subsection{DeepFake Creation and Detection}

GANs \cite{goodfellow2014generative} have achieved remarkable progress in image synthesis \cite{zhu2017unpaired} and voice synthesis \cite{oord2016wavenet}, which are widely employed in creating realistic DeepFakes. In this paper, we mainly focus on image synthesis which plays a key role in creating modern DeepFakes. Entire synthesis and partial synthesis are two typical manipulations in facial image synthesis with GANs \cite{tolosana2020deepfakes}. In the entire synthesis, the whole synthesized images are totally generated by GANs and it can be used for synthesizing a new face that does not exist in the world. PGGAN \cite{karras2017progressive} and StyleGAN \cite{karras2019style} can generate high-resolution facial images to improve the quality of a given face. Specifically, StyleGAN has the capability to synthesize a non-existent face by utilizing the idea of style transfer. In the partial synthesis, the face attributes like hair, expression, are manipulated by GANs automatically. StarGAN \cite{choi2018stargan}, STGAN \cite{liu2019stgan}, and AttGAN \cite{he2019attgan} can edit the attributes in a fine-grained manner, for example, changing the hair color, wearing eyeglasses, turning the smiling expression into scared, \etc. Thus, determining whether a facial image is manipulated by GANs provides a straightforward idea for detecting DeepFake.

Due to the imperfection design of existing GANs, the manipulated images with GAN inevitably introduces various artifacts. Existing studies on identifying DeepFakes are mostly leveraging the artifacts as clues. The artifacts can be classified as observable-artifacts noticed by human eyes and invisible-artifacts learned by DNN-based classifiers \cite{wang2020fakespotter,zhang2019detecting,fx_ijcai20_fakespotter,acmmm20_deepsonar,acmmm20_deeprhythm,huang2020fakelocator}.

Lyu \etal proposed to spot DeepFake video by observing the lack of eye blinking in the synthesized face \cite{li2018ictu}. The inconsistent head poses in the synthesized face is another observable-artifacts in DeepFake videos \cite{yang2019exposing}. Some researchers also investigated the invisible-artifacts which could be used for spotting DeepFakes. Wang \etal observed that CNN-generated images contain common artifacts that could be identified by careful pre- and post-processing and data augmentation \cite{wang2020cnn}. Frank \etal addressed the GAN-generated image identification with a basic observation that the artifacts revealed in the frequency domain \cite{frank2020leveraging}. AutoGAN \cite{zhang2019detecting} observed the upsampling design in GAN will introduce artifacts in the synthesized images, thus they developed a GAN simulator to produce fake images and train a classifier to detect GAN-generated images. These proposed methods all claimed the effectiveness on seen GANs, but their capabilities on unknown GANs are still unclear.

\subsection{DeepFake Disruption and Evasion}

Instead of detecting DeepFakes passively, some studies are working on disrupting the DeepFake creation proactively by adding adversarial noises into the input facial images.

Segalis \etal{} \cite{segalis2020disrupting} introduced spatial-temporal distortions to disrupt face-swapping manipulations by injecting minute perturbations to source video frames. Ruiz \etal{} \cite{ruiz2020disrupting} focus on the white-box and gray-box settings in DeepFake generation by presenting a spread-spectrum disruption on conditional image translation networks, rather than the simple evaluation on face-swapping manipulations in the aforementioned study \cite{segalis2020disrupting}. Chin-Yuan \etal{} \cite{yeh2020disrupting} introduced two types of adversarial attack (\eg{}, nullifying attack and distorting attack) on image-to-image translation models to output broken and disfigured images. Instead of employing the naive adversarial faces, Yang \etal{} \cite{yang2020defending} proposed to apply a novel transformation-aware adversarially perturbed faces to disrupt the DeepFake creation. They leverage differentiable random image transformations for generating perturbed faces, leading to synthesized faces with obvious visual artifacts.

These methods are all inspired by the adversarial faces which could result an erroneous GAN output. However, the perturbed faces could be easily detected by the existing  adversarial attack detection methods. Furthermore, these studies are all work in white-box or gray-box settings which need to obtain the knowledge of synthetic techniques. In contrast with these studies, our method could work in black-box setting and the embedded messages follow a stealthy manner.


On a similar note, some recent work aim at evading DeepFake detection through various image-level and frequency-level manipulations \cite{acmmm20_fakepolisher,huang2020fakeretouch,carlini2020evading,durall2020watch,jung2020spectral}. These work call for effective method for fighting against DeepFakes in a robust manner.

\subsection{Digital Watermarking}

In the past decades, digital watermarking plays a key role in digital multimedia copyright protection \cite{katzenbeisser2000digital,podilchuk2001digital,siddaraju2015digital}. The existing watermarking are mostly evaluated on various image transformations.


The spatial and frequency domain are two lines in embedding watermark into the carrier. Spatial domain is more easily to perform than the frequency domain, but it can be easily corrupted or attacked by attackers with pixel perturbations \cite{singh2012novel}. The spatial domain techniques embed watermark by modifying the pixels value, such as the least significant bit (LSB) \cite{bamatraf2010digital}. In embedding on the frequency domain, the carrier will be first converted into a specific transformation, then the watermark will be embedded in the transformation coefficients. The common frequency domains adopted in embedding watermarks include discrete cosine transform (DCT), discrete wavelet transform (DWT), discrete Fourier transform (DFT), and singular value decomposition (SVD) \cite{jiansheng2009digital,khan2013digital,yavuz2007improved}.

With the rapid development of deep learning, end-to-end watermark embedding techniques are proposed in recent years. HiDDeN \cite{zhu2018hidden} proposed the first end-to-end framework by jointly training encoder and decoder network which could robust to noises like Gaussian blurring, pixel-wise dropout, \etc. StegaStamp \cite{tancik2020stegastamp} presented a steganographic algorithm for embedding arbitrary hyperlink into the photos, which comprises a deep neural network for encoding and decoding.

\section{Problem Statement}\label{sec:pro}

In this paper, our real world system is described in Fig.~\ref{fig1}. A user could upload his/ her personal photos to social networks like Facebook and share it with friends or anyone. Unfortunately, attackers can easily pick victim's photos and manipulate them with various GANs to create DeepFakes they wanted, like releasing a fake statement in a video. The created DeepFakes will cause panic and raise security and privacy concerns for victims when it spreads on social networks. Our proposed FakeTagger embeds message into the images before uploading to the social networks, after which it tries to recover the embedded message from a suspicious photo in social network for DeepFake detection and DeepFake provenance by determining the sources based on the recovered message. The key idea here is that our image tagging method should be robust enough to survive the drastic image transformation and reconstruction by the DeepFake process. Finally, the confirmed DeepFakes could be blocked and avoid further spreading.

Here are more details regarding Fig.~\ref{fig1}. In the top panel, after a user (Fig.~\ref{fig1}-a1) uploads his/her personal photos to the public domain social media platform, the personal picture can be picked up by a malicious actor (Fig.~\ref{fig1}-b1). The bad actor can apply off-the-shelf DeepFake technology to produce a DeepFaked version of the user's face image (Fig.~\ref{fig1}-c1). In this case, the male face is transformed to exhibit female's attribute, which is one example of how DeepFake can alter any face image without noticeable artifacts. Then, the bad actor can upload the DeepFaked face image to the same social media platform again (Fig.~\ref{fig1}-d1), impersonating the user, or aiming at other malicious activities such as spreading misinformation. As can be seen, the unprotected social media platform is quite vulnerable in this scenario in terms of identifying the DeepFake images and preventing the spread of misinformation since no mechanism is established to distinguish between a legitimate face image and a DeepFake one. 

On the contrary, in the bottom panel where the social media platform is protected by the proposed FakeTagger mechanism, the spread of misinformation can be effectively prohibited. When a user uploads his/her personal photo (Fig.~\ref{fig1}-a2) to the social media platform, the FakeTagger is invoked to check whether this picture has been tagged by a FakeTagger message before (usually a UID that matches the user's identity). If this face image is new, FakeTagger can embed a message in the image, which is sufficiently robust to survive drastic image transformation such as DeepFake reconstruction. When a malicious bad actor (Fig.~\ref{fig1}-b2) picks out the victim's photo and applies the DeepFake technique (Fig.~\ref{fig1}-c2), the FakeTagger message will survive. Then, when the bad actor tries to upload the DeepFaked face image to the social media platform again (Fig.~\ref{fig1}-d2), the embedded FakeTagger message will trigger an alarm since the UID of the original picture does not match the one of the bad actors, indicating a perpetrating event has happened. In this way, proper measures can be taken to stop the spread of misinformation such as blocking the uploading of the DeepFake face image, and/ or raising a red flag for this bad actor. In the bottom panel, the FakeTagger protected images are represented by a green tag as well as a blue picture frame. In both panels, the pink arrows depict the route that a bad actor can take from picking a victim to the spread of misinformation. The blue arrow route indicates where  FakeTagger message remains active during the whole process.

\section{Methodology}\label{sec:method}


\subsection{Insight} \label{subsec:insight}

Existing techniques against DeepFake aim at observing the artifacts in the synthesized images with various methods. However, these studies suffer two issues, 1) they are not general to unknown GANs \cite{karras2020analyzing}, 2) they are easily susceptible to adversarial attacks by adding perturbations intentionally or simple image transformation (\eg, compression, Gaussian noises) \cite{qian2020thinking,carlini2020evading}. Thus, the existing artifact-based techniques are not prepared in tackling the future emerging DeepFake threats.


A straightforward idea for defending DeepFakes could be fighting them proactively. Disrupting the DeepFake creation and tracking the source of DeepFake by embedding tag in advance might be promising solutions. However, DeepFake disruption with adversarial faces which is fragile and could be easily discovered by the existing techniques. Alternatively, we explore whether a robust image tagging can be served as a safeguard for protecting the safety of facial images in social networks against DeepFake. The image tagging allows us to easily conduct DeepFake detection and provenance with the embedded message. A practical image tagging should satisfy the following properties:


\begin{itemize}[leftmargin=*]
\item Image tagging for DeepFake should be robust against GAN-based transformation, rather than simple image transformation like conventional digital watermarking.
\item The tagged message should be imperceptible to human eyes and without introducing obvious image quality decrease. In other words, the message need follow stealthiness property.
\end{itemize}

Inspired by the advances of deep learning in achieving significant progress in various computer vision problem, we employ a DNN based encoder and decoder and jointly trained for message embedding and recovering. Due to the introduced drastic transformation in DeepFake creation, we are motivated by the Shannon's capacity theorem that redundancy could improve robustness in signal communication. In our FakeTagger, we employ channel coding by injecting redundant message to improve the possibility in recovering messages after DeepFake manipulation. In the subsections, we introduce the pipeline of our proposed image tagging for DeepFake provenance tracking.

\subsection{Image Tagging Pipeline} \label{subsec:pipline}

\subsubsection{Overview}

\begin{figure*}[t!]
\centering
\includegraphics[width=0.80\linewidth]{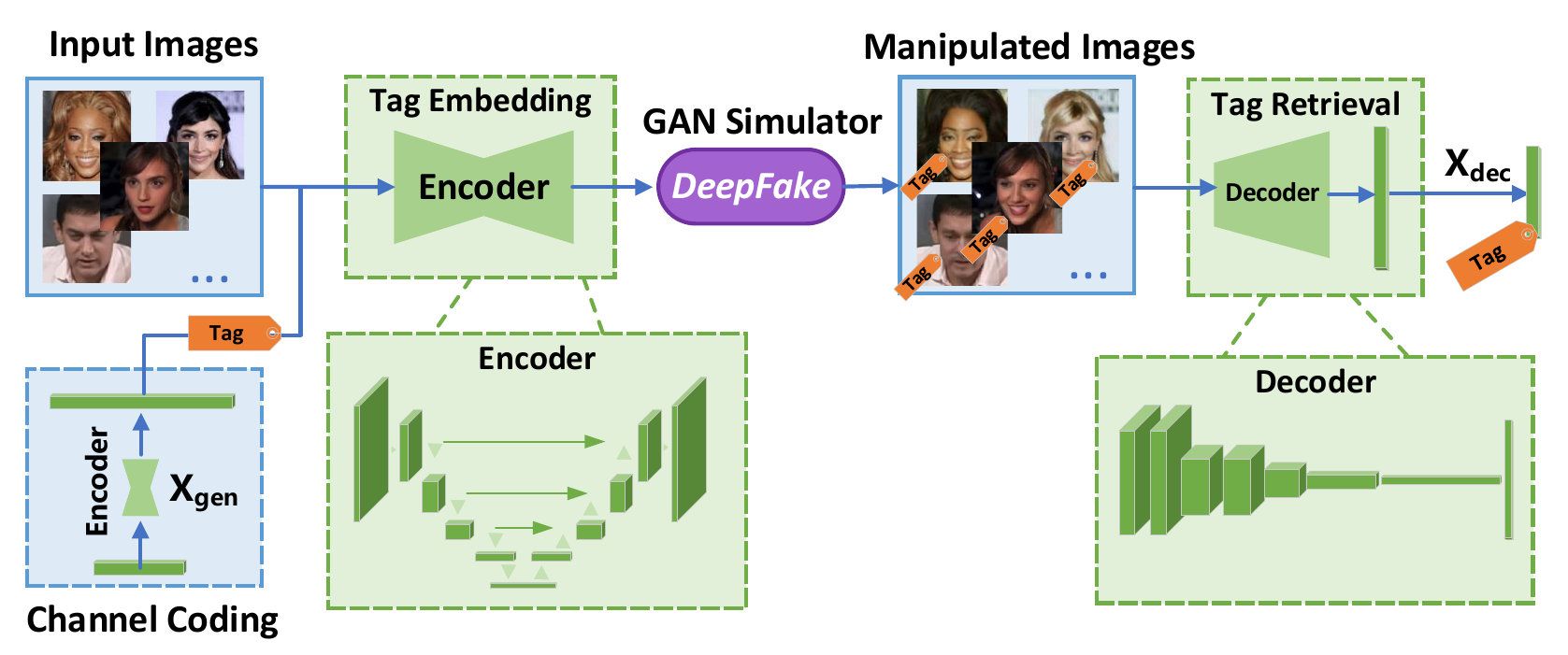}
\vspace{-10pt}
\caption{Overview of our proposed FakeTagger. The message generator $\mathnormal{X}_{gen}$ first generates a redundant message from the given input message $\mathnormal{X}$, then the encoder $\mathnormal{F}_{enc}$ encodes the input image $\bm{I}$ and redundant message $\mathnormal{X}^\prime$ to produce an encoded image $\bm{\widetilde{I}}$. The attack would manipulate the encoded image $\bm{\widetilde{I}}$ to generate various DeepFaked image. The decoder $\mathnormal{F}_{dec}$ recovers the message from the DeepFaked image and output message $\bm{\widetilde{X}}$. Finally, the channel decoder $\mathnormal{X}_{dec}$ accepts $\bm{\widetilde{X}}$ to produce the final message.}
\label{fig2}
\vspace{-10pt}
\end{figure*}

Fig.~\ref{fig2} gives an overview of our proposed FakeTagger overall architecture. Our method includes five key components, a message generator $\mathnormal{X}_{gen}$, a DNN-based encoder $\mathnormal{F}_{enc}$, a GAN simulator $\mathnormal{G}_{sim}$, a DNN-based message decoder $\mathnormal{F}_{dec}$, and a channel decoder $\mathnormal{X}_{dec}$. Specifically, the functionalities of each component as follows.

\begin{itemize}[leftmargin=*]
\item The message generator $\mathnormal{X}_{gen}$ generates binary message from channel coding. The generated message serves as an asset for identity verification.
\item The encoder $\mathnormal{F}_{enc}$ embeds a message (usually a UID) into a facial image and ensures the tagged message invisible to human eyes. In other words, the encoded image needs to be perceptually similar to the input image. 
\item The GAN simulator $\mathnormal{G}_{sim}$ is adopted for performing various GAN-based transformation, including identity swap, attribute editing, face reenactment, and entire synthesis.
\item The message decoder $\mathnormal{F}_{dec}$ recovers the embedded message from the encoded facial images after drastic GAN-based transformation. The recovered UID is further used for the identity verification purpose.
\item The channel decoder $\mathnormal{X}_{dec}$ accepts the decoded message from $\mathnormal{F}_{dec}$ to produce the final message $\mathnormal{X}$.
\end{itemize}

\subsubsection{Image tagging encoder-decoder training} 
The DNN-based encoder and decoder are jointly trained to embed messages into the given input facial images. The encoder allows an arbitrary message to imperceptibly embed into the given arbitrary facial images. The decoder is trained to retrieve the embedded message even after drastic GAN-based manipulation, like partial attribute editing. Here, the embedded message indicates $n$ bits UID, but it can be easily extended to arbitrary binary bits.

Specifically, the encoder $\mathnormal{F}_{enc}$ receives a facial image $\bm{I}$ and a message $\bm{X}$ as input, then the message generator $\mathnormal{X}_{gen}$ produces a redundant message $\mathnormal{X}^\prime$. The encoder $\mathnormal{F}_{enc}$ outputs a tagged facial image $\bm{\widetilde{I}}$ with a mapping $\mathnormal{F}_{enc}(\bm{I},\mathnormal{X}^\prime)\mapsto\bm{\widetilde{I}}$. The input facial image $\bm{I}$ need to perceptually similar to the encoded facial image $\bm{\widetilde{I}}$, where $\bm{I} \thickapprox \bm{\widetilde{I}}$. The encoded facial images may manipulated by GAN, where $\mathnormal{G}_{sim}(\bm{\widetilde{I}})\mapsto\bm{\overline{I}}$. The decoder try to recover the embedded message $\mathnormal{F}_{dec}(\bm{\overline{I}})\mapsto\bm{\widetilde{X}}$ or $\mathnormal{F}_{dec}(\bm{\widetilde{I}})\mapsto\bm{\widetilde{X}}$, where $\bm{\widetilde{X}} \thickapprox \bm{X}^\prime$. Finally, the channel decoder $\mathnormal{X}_{dec}$ produces the final message $\bm{X}$.


\subsubsection{GAN-based manipulation} 

DeepFake involves four types of facial images manipulation with various GANs. Specifically, they are all the existing DeepFake types, namely identity swap, face reenactment, attribute editing, and entire synthesis. In the real world scenario, our encoded facial images $\bm{\widetilde{I}}$ will be manipulated by these four types of GAN-based manipulations. Thus, a GAN simulator performs the two typical manipulations by connecting our encoder and decoder to enforce that the decoder could learn how to recover message after drastic GAN-based manipulations.

\subsubsection{Channel coding} In signal transformation, channel coding is applied for correcting errors \cite{bossert1999channel}. Specifically, channel coding is designed for addressing the limitation of data transferring in a noisy channel. Here, we apply channel coding for injecting redundant message to hope that our embedded message could survive the drastic GAN-based transformation. In our work, the GAN-based transformation can be simply deem as a kind of noisy channel.

\begin{figure}[t!]
\centering
\includegraphics[width=0.99\linewidth]{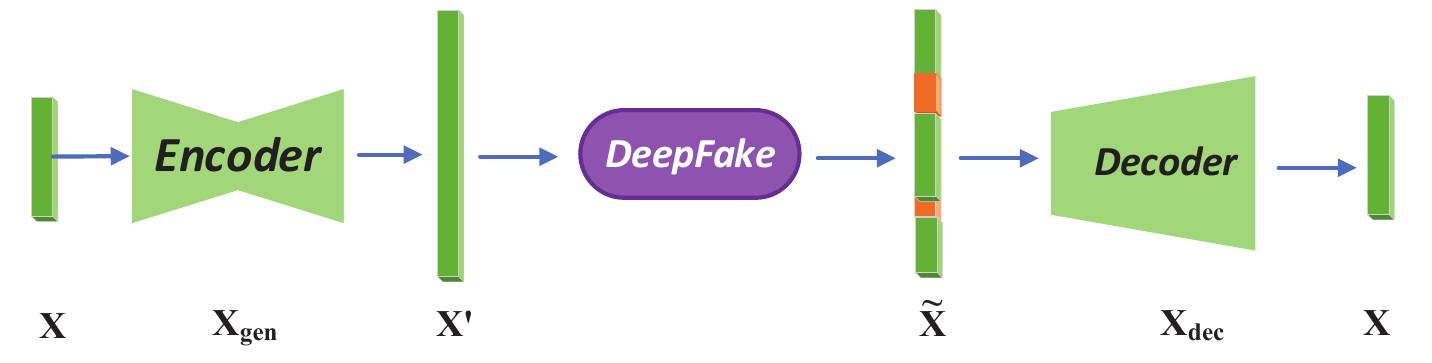}
\vspace{-10pt}
\caption{Overview of channel coding.}
\label{fig_channel}
\vspace{-10pt}
\end{figure}

Figure \ref{fig_channel} illustrates our adopted channel coding. Given a binary message $\mathnormal{X}\in\{0,1\}^L$ of length $L$. Our message generator $\mathnormal{X}_{gen}$ produces a redundant message $\mathnormal{X}^\prime$ where the length is large than $L$. In this work, the channel distortions is the errors introduced by the GAN-based transformation. We apply a binary symmetric channel (BSC) to formulate the channel distortion. BSC is a standard channel distortion model that assumes each bits is in the message independently and randomly flipped with a probability $p$. In our experiments, we find that BSC works well for our work. It will be interesting to explore other distortion models which could perfect formulate our problem.

\subsubsection{Losses}

To enforce minimal error of decoded message, we introduce the two major losses in the model training. The message loss $\mathcal{L}_C$ calculates the loss between the decoded message and the generated message with $\mathnormal{X}_{gen}$. Here, we use $L_2$ loss. The message loss is represented as $\mathcal{L}_C=\lambda \left \| \bm{\widetilde{X}} - X^\prime \right\|^2$. The image loss $\mathcal{L}_M$ measures the similarity between encoded image and the input image. We use $L_2$ loss and a GAN loss $L_G$ loss with spectral normalization \cite{miyato2018spectral} to preserve the visual quality of encoded image. The image loss is represented as $\mathcal{L}_M=\alpha \left \|I- \bm{\widetilde{I}} \right\|^2 + \beta L_G(\bm{\widetilde{I}})$. The training loss is the weighted sum of these loss components.


\section{Experimental Setting and Implementation}\label{sec:exp}


\subsection{Data Preparation}\label{sec:data}

\subsubsection{GANs} In our experiments, we employ DeepFaceLab \cite{petrov2020deepfacelab} for identity swap, Face2Face \cite{thies2016face2face} for face reenactment, STGAN \cite{liu2019stgan} for attribute editing, and StyleGAN \cite{karras2019style} for entire synthesis, since they achieved the state-of-the-art performance in DeepFake creation. DeepFaceLab provides a framework for face swapping, Face2Face can transfer facial expression to target image, STGAN edits facial attributes (\eg, wearing eyeglasses, changing hair color) in a fine-grained manner, StyleGAN can reconstruct a given face and generate a new face.


\subsubsection{Dataset} We employ CelebA-HQ \cite{karras2017progressive} that is a public face dataset consisting 30,000 facial images and contains several different size facial images, such as $128\times128$, $512\times512$, and $1,024\times1,024$, \etc. In our experiments, we explore the effectiveness of FakeTagger in tackling facial images with different input size.

\subsection{Baseline} 

In evaluation, a straightforward idea to demonstrate the effectiveness of our approach is compare with the conventional digital watermarking techniques as baseline like spatial-domain and frequency-domain watermarking. However, in our initial experiments employing LSB and DWT for digital watermarking, both of them are all failed in recovering messages after the four types of DeepFake manipulation. Thus, in our experiments, the baseline is a deep learning based embedding technique without introducing redundant message injection.

\subsection{Evaluation Metrics}

To evaluate the performance of FakeTagger quantitatively, we employ accuracy to measure the recovered message after GAN-based manipulations. The accuracy indicates the full message retrieval rate (FMRR). PSNR and SSIM are adopted for calculating the similarity between the input and encoded facial images with FakeTagger.

\subsection{Implementation}

\subsubsection{Encoder} Our encoder is trained to embed messages into carrier images while preserving the perceptual similar to the input carrier. Here, we use a U-Net \cite{ronneberger2015u} style architecture for receiving the input carrier images and output an encoded three-channel image. In our experiments, we explore different size of input carrier images (\eg, $128\times128$, $512\times512$) and different length of embedded message (\eg, $20$ bits, $30$ bits, $50$ bits). Furthermore, the embedded message could be embedded in different levels in our encoder for achieving better performance in recovering the message in the decoder.

\subsubsection{Decoder} Our decoder is trained to retrieve the embedded message from the encoded images that are the output of our encoder. 
The decoder consists of seven convolutional layers with kernel size $3\times3$ and strides $\ge1$, one dense layer, and finally output the decoded message with the sigmoid activation function. The size of the decoded message is the same as the embedded message.

\subsubsection{GAN simulator} In the white-box setting, we directly apply DeepFaceLab, Face2Face, STGAN, and StyleGAN serving as the GAN simulator. In the black-box setting, we employ a GAN simulator proposed in AutoGAN \cite{zhang2019detecting} to simulate the generation of DeepFake transformation. More details refer to Section \ref{Sec:result}.


\subsubsection{Channel coding} In general, any standard error correcting code like low-density parity-check (LDPC) codes \cite{ryan2004introduction} for generating $X^\prime$ in the message generator $\mathnormal{X}_{gen}$. However, LDPC need an estimation of the length of noises which is not practical. Here, we use NECST \cite{choi2018necst} for source and channel coding with a learning model, which has no restriction on the noise length. Specifically, BSC is adopted for training the channel distortion model and the input $\mathnormal{X}$ is randomly sampled. Specifically, our channel coding model is not jointly trained with the $\mathnormal{F}_{enc}$ and $\mathnormal{F}_{dec}$ to avoid co-adaption with specific DeepFake manipulations, which results in overfitting.

\subsubsection{Encoder and decoder training} The encoder and decoder are jointly trained with randomly generated messages. The input images are collected from the public dataset CelebA-HQ. In training, we use $4$ different sizes input facial images to train the model to explore the performance of FakeTagger in tackling input faces of different sizes.

\section{Experimental Results}

In experiments, our evaluation aims to answer the following three research questions.

\begin{itemize}[leftmargin=*]
    \item \textbf{RQ1}: What is the performance of FakeTagger in recovering the embedded messages in white-box and black-box settings with different DeepFake manipulation.
    \item \textbf{RQ2}: Whether our FakeTagger is robust against the common image transformation and perturbations, such as compression, resizing, \etc.
    \item \textbf{RQ3}: Whether the encoded image with embedded messages is stealthiness to human eyes and preserve a good visual quality.
\end{itemize}

\subsection{Effectiveness Results (RQ1)} \label{Sec:result}

In this section, we mainly explore the effectiveness of our proposed FakeTagger in recovering the embedded message with four types of DeepFakes. Four typical DeepFakes are adopted in our experiments for evaluation, namely DeepFaceLab for identity swap, Face2Face for face reenactment, STGAN for attribute editing, and StyleGAN for entire synthesis are adopted for evaluation. Here, the length of the message is set to $30$ bits, the redundant message size is $150$ bits. More experiments on exploring the  performance of redundant message size is presented in our ablation study in Section \ref{Sec:ablation}. Additionally, we conduct extensive experiments to illustrate whether the introduced channel coding could help to improve the performance.


\textbf{Effectiveness on White-box}. Tab.~\ref{Table:eff_1} summarizes the performance of FakeTagger in tackling with the three types of DeepFakes. Experimental results shown that our FakeTagger performs well in the identity swap and face reenactment which could be consider as partial synthesis. The best result gives an accuracy 97.3\% and the worst result gives an accuracy 95.7\%. However, the best result of FakeTagger in entire synthesis is 95.2\%.  The main reason is that the our FakeTagger is susceptible to the manipulation region and entire synthesis involves more drastic manipulation than the partial synthesis like identity swap and face reenactment. We also observe that the size of the input image has a positive impact on performance. Large size image can provide more space for embedding message and can survive in GAN-based manipulation more easily. Furthermore, the experimental results also tell us that injecting redundant message can significantly improve the performance of our FakeTagger in surviving various DeepFake manipulation.

Tab.~\ref{Table:eff_2} presents the performance of FakeTagger in dealing with attribute editing by employing STGAN. The manipulated attributes include removing hair into bald, adding mustache, wearing eyeglasses, and changing into pale skin. Experimental results have shown that our FakeTagger can perform well in the three former attributes manipulation, but susceptible to the skin color changing. The main reason is also that the manipulation region is larger and the intensity is drastic than other three attribute editing.

The experimental results in Tab.~\ref{Table:eff_1} and Tab.~\ref{Table:eff_2} show that our FakeTagger achieves an average accuracy more than 95\% in message recovering over the four types of DeepFake. These two tables also tell us that our FakeTagger is sensitive to the region of manipulation and the input image size. Additionally, the injected redundant messages play a key role in improving the performance of our FakeTagger, large input image size leading to better performance in  message recovering.

\begin{table}[]
\scriptsize
\centering
\caption{Performance (FMRR) of FakeTagger on three types of DeepFakes in white-box setting. R indicates that the message inject redundant messages with channel coding, N denotes that the message without injecting any redundant messages.}
\vspace{-10pt}
\setlength{\tabcolsep}{3.5pt}
\begin{tabular}{c|c|c|c|c|c|c}
\toprule
\multirow{2}{*}{\textbf{Image Size}} & \multicolumn{2}{c|}{\textbf{Identity Swap}} & \multicolumn{2}{c|}{\textbf{Face Reenactment}} & \multicolumn{2}{c}{\textbf{Entire Synthesis}}\\
&\textbf{R}& \textbf{N}&\textbf{R}& \textbf{N}&\textbf{R} & \textbf{N}\\ \midrule
$128\times128$ & 0.963 & 0.837 & 0.957 & 0.820 & 0.928 & 0.736 \\ 
$256\times256$ & 0.969 & 0.840 & 0.961  & 0.831 & 0.933 & 0.749 \\ 
$512\times512$ & 0.973 & 0.859 & 0.968   & 0.833 & 0.952 & 0.780 \\ 
Average & \textbf{0.968} & 0.845 & \textbf{0.962} & 0.828  & \textbf{0.938} & 0.755 \\ 
\bottomrule
\end{tabular}
\label{Table:eff_1}
\vspace{-5pt}
\end{table}

\begin{table}[]
\scriptsize
\centering
\caption{Performance (FMRR) of FakeTagger on attribute editing types of DeepFakes in white-box setting. R indicates that the message inject redundant messages with channel coding, N denotes that the message without injecting any redundant messages. Manipulating the color of skin is the most drastic one.}
\vspace{-10pt}
\setlength{\tabcolsep}{3.5pt}
\begin{tabular}{c|c|c|c|c|c|c|c|c}
\toprule
\multirow{2}{*}{\textbf{Image Size}} & \multicolumn{2}{c|}{\textbf{bald}} & \multicolumn{2}{c|}{\textbf{mustache}} & \multicolumn{2}{c|}{\textbf{eyeglasses}} & \multicolumn{2}{c}{\textbf{plain skin}}\\
&\textbf{R}& \textbf{N}&\textbf{R}& \textbf{N}&\textbf{R} & \textbf{N}&\textbf{R} & \textbf{N}\\ \midrule
$128\times128$ & 0.975 & 0.849 & 0.983 & 0.850 & 0.971 & 0.852 & 0.968 & 0.842 \\ 
$256\times256$ & 0.981 & 0.852 & 0.988  & 0.850 & 0.973 & 0.855 & 0.969 & 0.847 \\ 
$512\times512$ & 0.983 & 0.856 & 0.991   & 0.861 & 0.978 & 0.861 & 0.973 & 0.848 \\ 
Average & \textbf{0.980} & 0.885 & \textbf{0.987} & 0.854 & \textbf{0.974} & 0.856 & \textbf{0.970} & 0.846 \\ 
\bottomrule
\end{tabular}
\label{Table:eff_2}
\vspace{-5pt}
\end{table}


\textbf{Effectiveness on Black-box}. Tab.~\ref{Table:eff_black} presents the performance of FakeTagger in dealing with four types of DeepFakes in total black-box setting. Experimental results shown that FakeTagger gives an average accuracy more than 88.95\% on the four types of DeepFake to demonstrate the effectiveness of our method in black-box setting.

Here, the GAN simulator simulates the GAN generation pipeline and generates a simulated ``manipulated'' image, rather than employing the specific GAN models such as STGAN, StyleGAN for image manipulation. The simulator is from AutoGAN \cite{zhang2019detecting} containing a generator $\mathcal{G}$, and a discriminator $\mathcal{D}$ with $l_1$ norm loss. In the generator, the decoder contains up-sampling module such as nearest neighbor interpolation with a general GAN architecture. The output of the generator try to reconstruct the original image which is similar to the original. Thus, our GAN simulator is more like an entire synthesis.

According to our experimental results, the baseline reaches an accuracy less than 60\% in message retrieval in black-box setting without obtaining any knowledge of the DeepFake techniques. In the black-box setting, FakeTagger performs well in the entire synthesis in compared with the other three DeepFakes. The main reason is that our GAN simulator is more like entire synthesis by reconstructing the input images. It would be more interesting to explore other GAN simulators like conducting fine-grained attribute editing, which would be our future work.

\begin{table}[]
\scriptsize
\centering
\caption{Performance (FMRR) of FakeTagger on four types of DeepFakes in black-box setting where the knowledge of DeepFake manipulation is unknown. In attribute editing, the manipulated facial attribute is changing the color of skin which is the most drastic facial attribute manipulation.}
\vspace{-10pt}
\setlength{\tabcolsep}{2.8pt}
\begin{tabular}{c|c|c|c|c}
\toprule
\textbf{Image Size} & \textbf{Identity Swap} & \textbf{Face Reenactment} & \textbf{Entire Synthesis} & \textbf{Attribute Editing}\\ \midrule
$128\times128$ & 0.857  & 0.872  & 0.901  & 0.883  \\ 
$256\times256$ & 0.878  & 0.877   & 0.912  & 0.889  \\ 
$512\times512$ & 0.895  & 0.891   & 0.920  & 0.897  \\ 
Average & \textbf{0.877}  & \textbf{0.880}  & \textbf{0.911}  & \textbf{0.890} \\ 
\bottomrule
\end{tabular}
\label{Table:eff_black}
\vspace{-5pt}
\end{table}

In summary, experimental results demonstrate the effectiveness of our FakeTagger in both white-box and black-box settings for message retrieval across the four types of DeepFake transformation. The experimental results also tell us that a large input image size and less region manipulation has a positive impact on the performance. Among the four types of DeepFake, entire synthesis would be more changing due to the drastic manipulation introduced in compared with the other three DeepFakes. 

\subsection{Evaluation on Robustness (RQ2)}

In creating DeepFake videos, the manipulated images will be further processed by numerous image perturbations like compression, resizing, \etc. In this section, we evaluated the robustness of FakeTagger in tackling these image perturbations which are common appeared in producing videos. 

\begin{figure}[tbp]
\centering
\subfigure[Compression]{
\includegraphics[width=0.45\columnwidth]{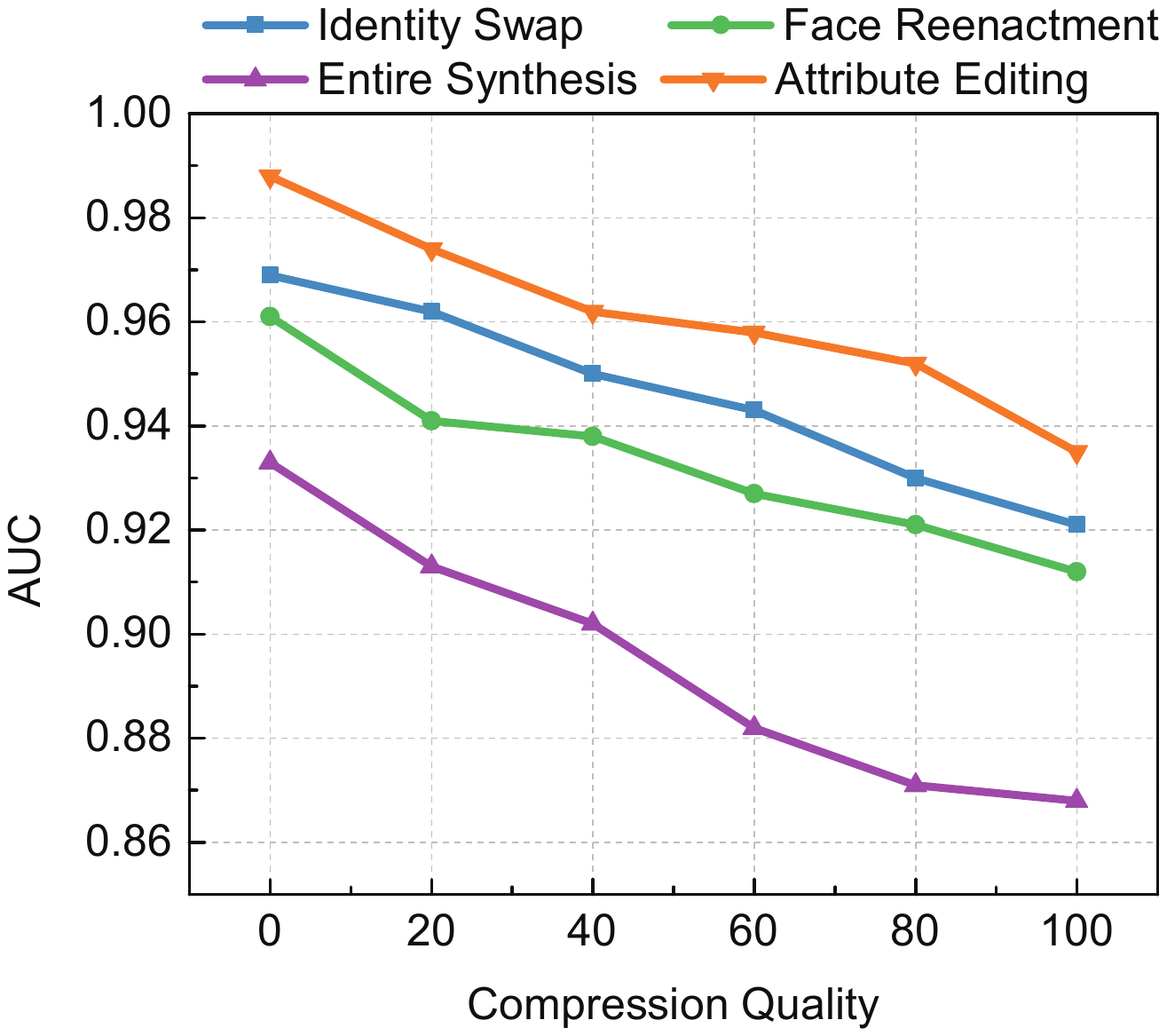}
}
\quad
\subfigure[Blur]{
\includegraphics[width=0.45\columnwidth]{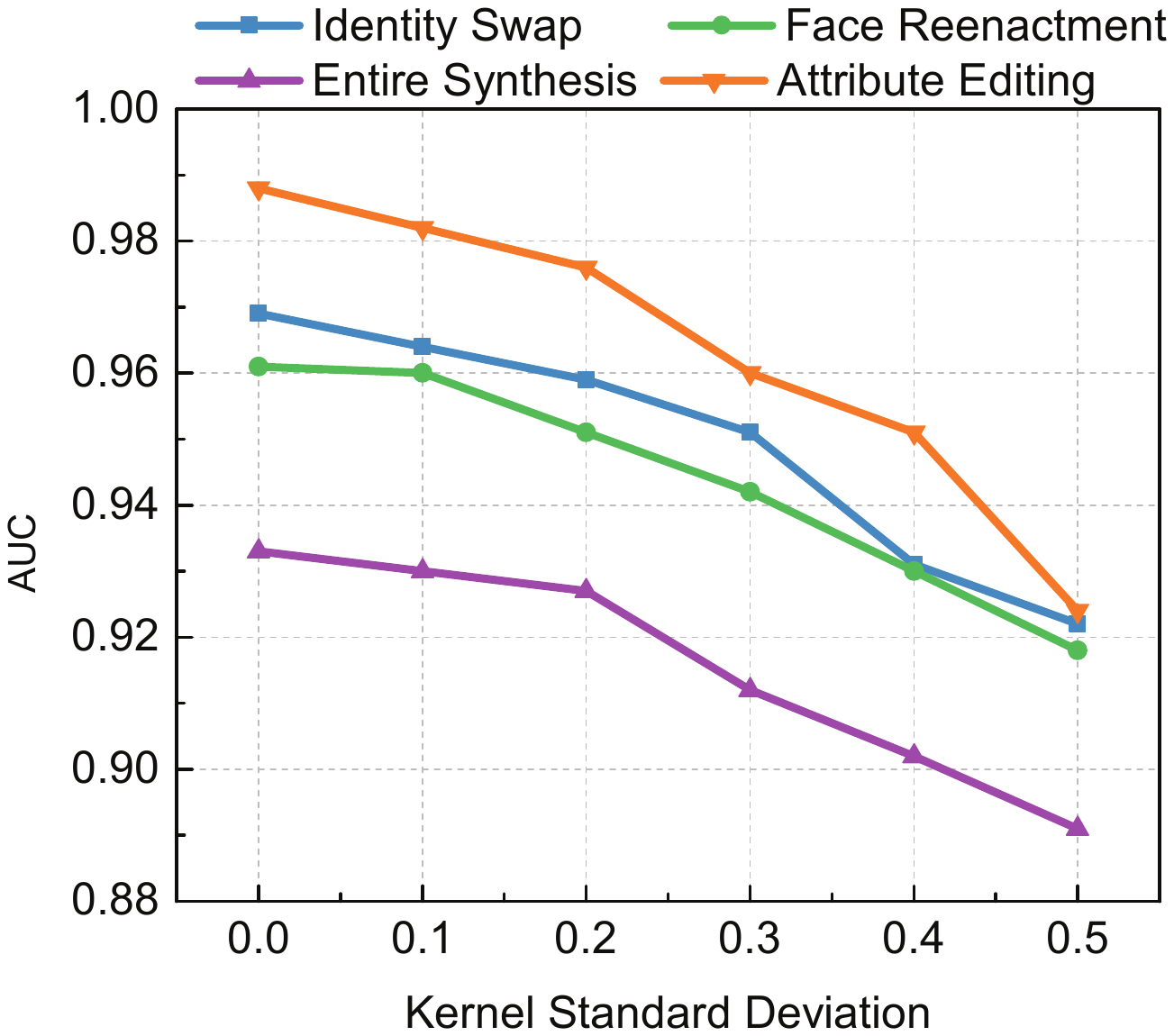}
}
\quad
\subfigure[Resizing]{
\includegraphics[width=0.45\columnwidth]{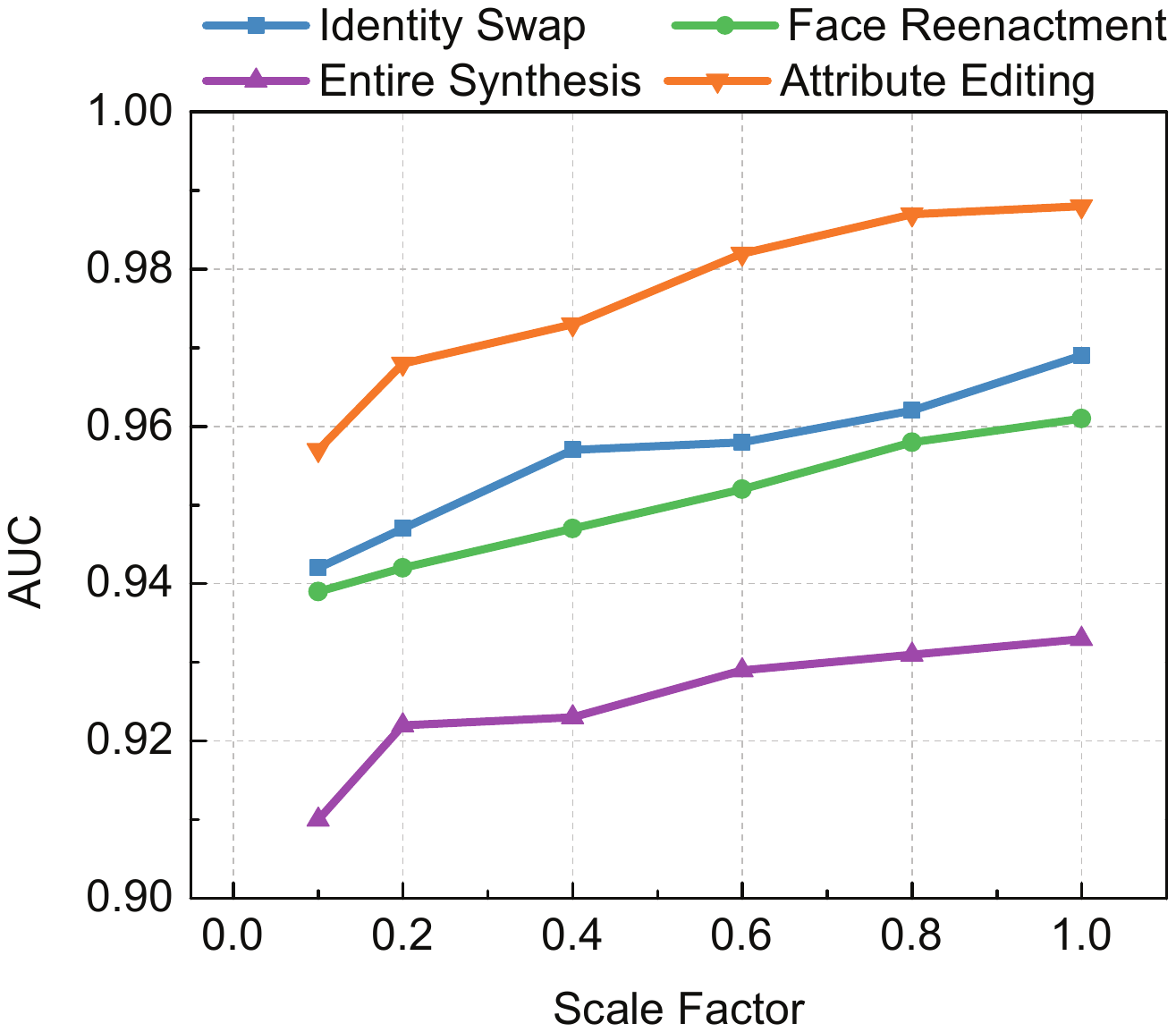}
}
\quad
\subfigure[Noise]{
\includegraphics[width=0.45\columnwidth]{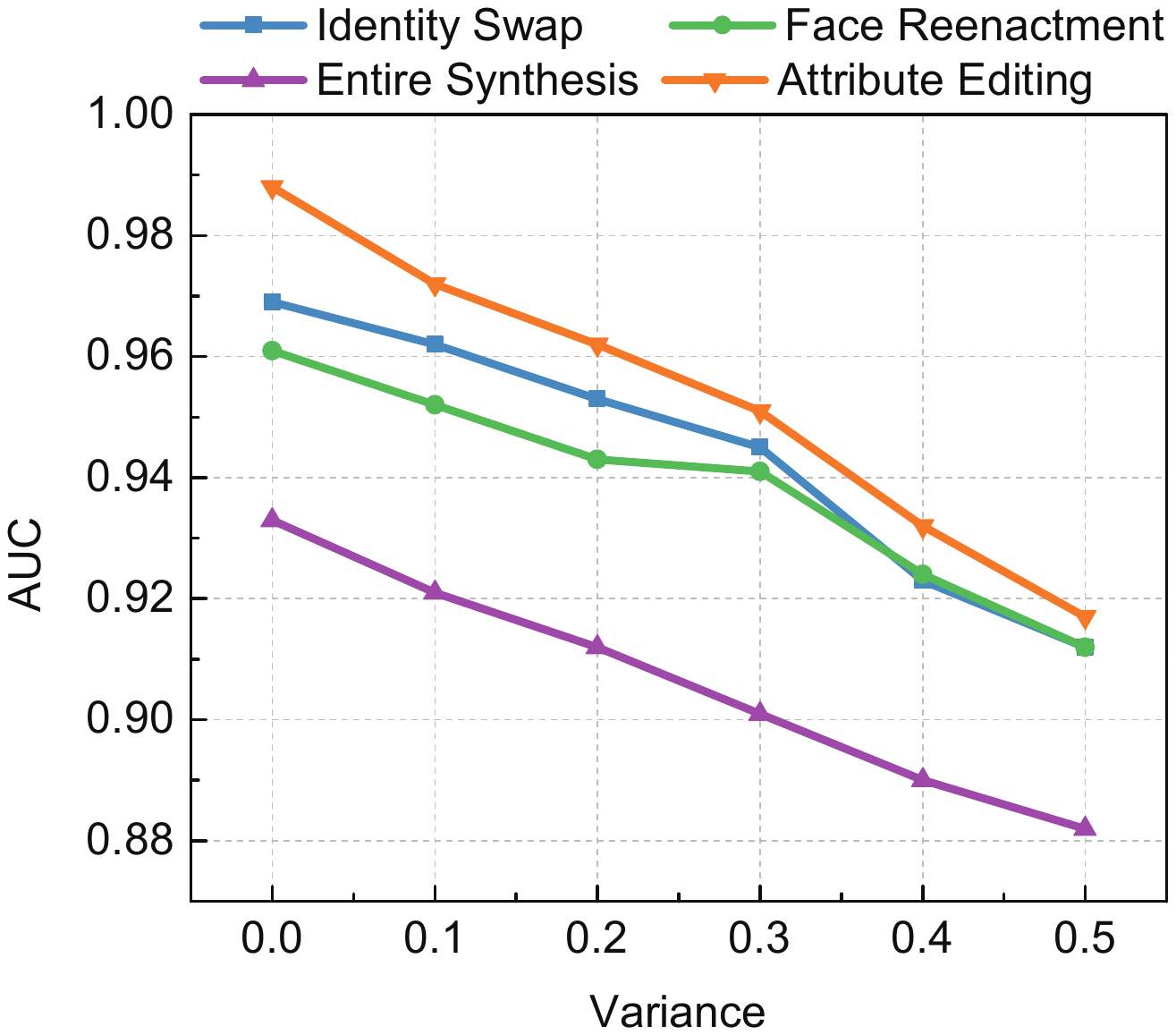}
}
\vspace{-10pt}
\caption{Robustness evaluation with four common degradations.}
\label{Figure:robusty}
\vspace{-17pt}
\end{figure}

Fig.~\ref{Figure:robusty} presents the robustness evaluation results of FakeTagger on four DeepFakes. In experiments, we employ four widely appeared perturbations in creating DeepFake videos, namely compression, resizing, blurring, and Gaussian noise. In experiments, the input image size is $256\times256$, and the manipulated facial attribute is ``mustache''. In  Fig.~\ref{Figure:robusty}, the compression quality measures the intensity of compression, range from 100 to 0. Blur means that the manipulated images are added with Gaussian blur. The kernel standard deviation is parameter for controlling the  intensity of blur. In experiments, the Gaussian kernel size to (3, 3). The scale factor in resizing is used for controlling the size of an image. The variance is used for control the intensity of added Gaussian noise.

Experimental results demonstrated the effectiveness of our FakeTagger in tackling with the four perturbations. We find that FakeTagger maintain a minor fluctuation range when the intensity of perturbation increases such as compression rate, the portion of resizing. Among the four types of DeepFakes, the entire synthesis is more sensitive to the four perturbation attacks, especially in the compression and adding Gaussian noises. Thus, our FakeTagger could be well applied for real application in considering the robustness against perturbations.

Our pioneering work leverages image tagging for defending DeepFakes proactively. In the performance evaluation, we consider the most strict case where all the bits are fully recovered. FakeTagger will have an even broader application, more robustness, and stronger resilience when partial errors could be tolerated in the message retrieval.

\subsection{Measuring the Stealthiness (RQ3)}

In FakeTagger, the encoder outputs an encoded image with embedded message. Ideally, the encoded images should be perceptually similar to the input image.We use two different metrics, PSNR and SSIM for measuring the distance between encoded image and input. 

Result in Tab.~\ref{Table:metrics} illustrates that our encoded image could maintain high visual quality. The PSNR value for the four DeepFakes are all large than 30dB, while the SSIM value for the four DeepFakes are lager than 0.93. According to the experimental results in Tab.~\ref{Table:metrics}, the entire synthesis achieved the best performance among the four DeepFakes, due to that the entire synthesis exhibits less artifact.

\begin{table}[]
\scriptsize
\centering
\caption{Image quality of the encoded images and input measured by PSNR and SSIM. For PSNR and SSIM, the higher the better.}
\vspace{-10pt}
\setlength{\tabcolsep}{2.5pt}
\begin{tabular}{c|c|c|c|c}
\toprule
\multirow{2}{*}{\textbf{Metrics}} & \multicolumn{4}{c}{\textbf{DeepFake}}\\ 
& \textbf{Identity Swap} & \textbf{Face Reenactment} & \textbf{Entire Synthesis} & \textbf{Attribute Editing} \\ \midrule
PSNR $\uparrow$ & 32.45 & 33.78  & 35.21 & 34.70 \\ 
SSIM $\uparrow$ & 0.931 & 0.939  & 0.948 & 0.942 \\
\bottomrule
\end{tabular}
\label{Table:metrics}
\vspace{-10pt}
\end{table}

\subsection{Ablation Study} \label{Sec:ablation}

In this section, we explore the impact of input and redundant message size on the performance of FakeTagger in recovering message. 

Capacity is an important factor for measuring the capability of our FakeTagger in embedding message. A large capacity indicates that the carrier can contain more information which could represent a large number of UID in our work. The success of our FakeTagger relies on introducing channel coding with redundant message which could tolerate a certain extent errors and recover the message correctly. Thus, in our work, we explore this two message size on the performance of FakeTagger.

\begin{SCfigure}
\centering
\includegraphics[width=0.63\linewidth]{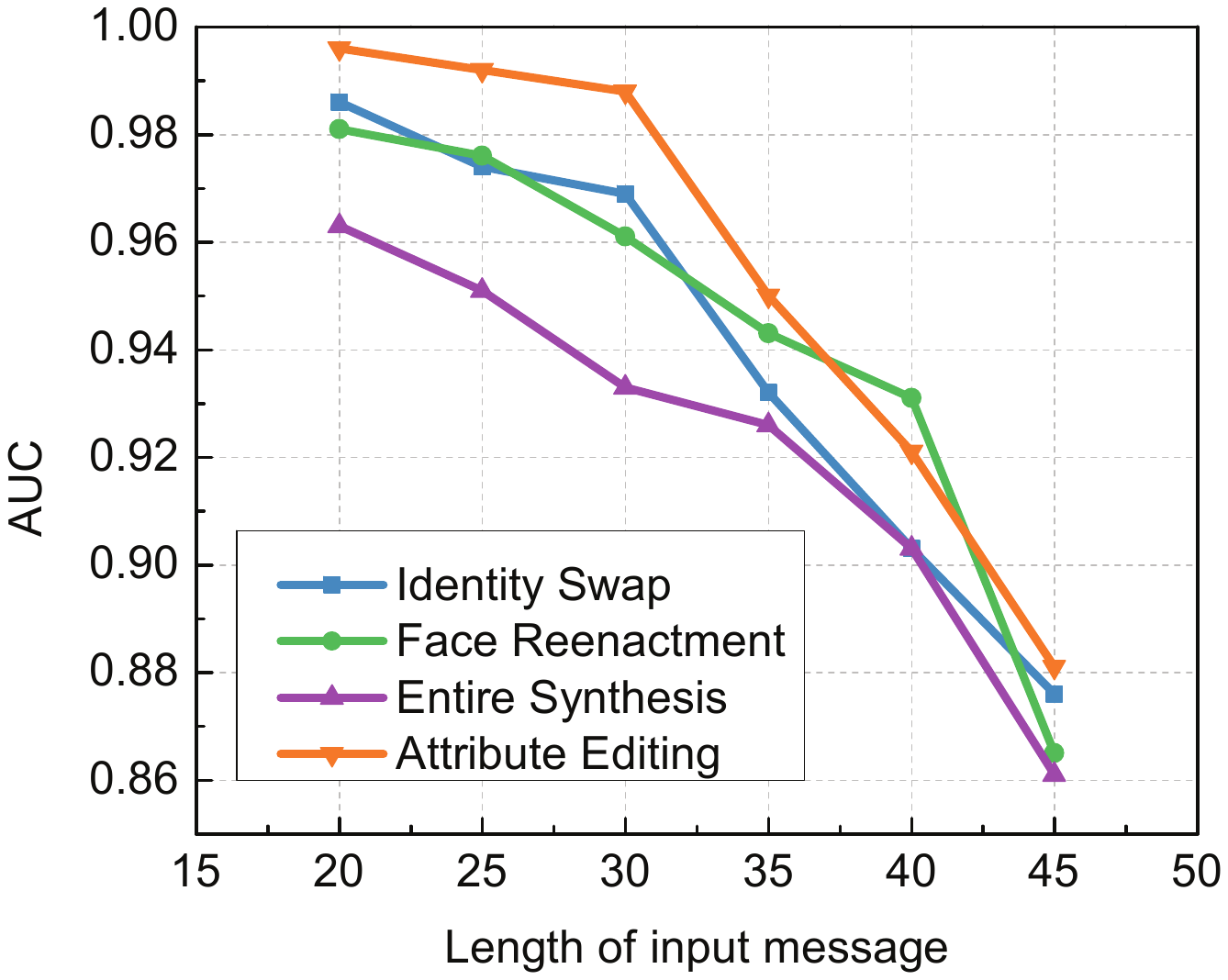}
\caption{Performance of FakeTagger on different size of input messages.}
\label{fig3}
\vspace{-10pt}
\end{SCfigure}

Fig.~\ref{fig3} shows the relation between the accuracy of FakeTagger in recovering messages and the length of message on four DeepFakes GANs. For the four types of DeepFakes, the input image size is 256$\times$256 which is the most common size in sharing images on the social networks. We select the mustache attribute with STGAN. The redundant message size is $150$ bits for all the input message. 

Results show that the length of message has a negative impact on the performance in recovering embedded message. FakeTagger can achieve an accuracy of more than 95\% on the four DeepFakes when the size of input message is $20$ bits and the redundancy rate is 750\%. However, the accuracy reduces to less than 90\% when the size of embedded message is $45$ bits and the redundancy rate is 333.3\%. Actually, the $30$ bits of message can represent more than $1$ billion different UIDs and the $35$ bits can represent more than $34$ billion UIDs, where the redundancy rate is 500\%, and 428\%, respectively. We believe that message with the $30$ bits or $35$ bits is enough for a social media platform to assign each user a specified UID. 

\subsection{Discussion}

FakeTagger achieves achieves competitive results in terms of both effectiveness, robustness and stealthiness on the four common DeepFakes, including identity swap, face reenactment, attribute editing, and entire synthesis.  However, FakeTagger also exhibits some limitations. First, in an adversarial environment, attackers could add adversarial noises to disrupt our embedded message for DeepFake provenance, and there is a tradeoff between generating imperceptible facial images and the success of disruption. Second, to survive various drastic DeepFake manipulation, especially the GAN-based transformation, our FakeTagger need to inject redundant messages, which introduces computation costs and the large redundant message has the potential to be observed by machine.

\section{Conclusion}\label{sec:concl}

In this paper, we proposed FakeTagger that embeds messages into the images for DeepFake provenance. To the best of our knowledge, this is the first work that presents a new insight for fighting against DeepFake from the perspective of privacy-preserving, which aims to defend DeepFake proactively. Experiments on four common DeepFakes including both entire synthesis and partial synthesis (\eg{}, identity swap, face reenactment, attribute editing) demonstrate the effectiveness, robustness, and stealthiness of our method in embedding messages into facial images and recovering them from facial images after drastic GAN-based transformation.

With the rapid development of AI-techniques, nobody can imagine future advances in producing DeepFakes. We can confirm that the DeepFake will become more and more realistic and everyone could fall victim. However, detecting DeepFakes by observing the artifacts in the synthesized images is obviously insufficient for protecting us against this AI risk. Our work poses a new insight for fighting against DeepFakes proactively, instead of observing the artifacts by leveraging domain knowledge in synthesized images which could easily become invalid in unseen GANs. 


Looking beyond DeepFake provenance tracking using the proposed FakeTagger, it is worth exploring if the FakeTagger can be used for the provenance tracking on other adversary modalities such as non-additive adversarial attacks ranging from adversarial weather elements such as rain \cite{arxiv20_advrain} and haze \cite{gao2021advhaze}, image degradation-mimetic adversarial attacks such as adversarial exposure \cite{icme21_xray,arxiv20_retinopathy}, vignetting \cite{ijcai21_ava}, blur \cite{neurips20_abba}, color jittering \cite{arxiv20_cosal}, \etc.

\begin{acks}
This research was supported in part by the fellowship of China National Postdoctoral Program for Innovative Talents No.BX2021229, the Fundamental Research Funds for the Central Universities No. 2042021kf1030, the National Natural Science Foundation of China (NSFC) under Grants No. 61876134, No. U1836112.
\end{acks}

\bibliographystyle{ACM-Reference-Format}
\bibliography{ref}

\end{document}